\documentclass[modern]{aastex61}
\usepackage{natbib}
\usepackage{xspace}
\pdfoutput=1
\shorttitle{Observing TESS Single Transit Candidates}
\shortauthors{Yao et al.}
\newcommand{\tess}{{\it TESS}\xspace}
\newcommand{\kepler}{{\it Kepler}\xspace}

\begin{document}

\title{Following up TESS Single Transits With Archival Photometry and Radial Velocities}

\author[0000-0003-4554-5592]{Xinyu Yao}
\affiliation{Department of Physics, Lehigh University, 16 Memorial Drive East, Bethlehem, PA 18015, USA}

\author[0000-0002-3827-8417]{Joshua Pepper}
\affiliation{Department of Physics, Lehigh University, 16 Memorial Drive East, Bethlehem, PA 18015, USA}

\author[0000-0003-0395-9869]{B.\ Scott Gaudi}
\affiliation{Department of Astronomy, The Ohio State University, 140 West 18th Avenue, Columbus, OH 43210, USA}

\author[0000-0002-4297-5506]{Paul A. Dalba}
\altaffiliation{NSF Astronomy \& Astrophysics Postdoctoral Fellow}
\affiliation{Department of Earth \& Planetary Sciences, University of California Riverside, 900 University Ave, Riverside, CA 92521, USA}

\author[0000-0002-0040-6815]{Jennifer A. Burt}
\affiliation{Jet Propulsion Laboratory, California Institute of Technology, 4800 Oak Grove Drive, Pasadena CA 91109, USA}

\author[0000-0001-9957-9304]{Robert A. Wittenmyer}
\affiliation{University of Southern Queensland, Centre for Astrophysics, West Street, Toowoomba, QLD 4350, Australia}

\author[0000-0003-2313-467X]{Diana~Dragomir}
\affiliation{Department of Physics and Astronomy, University of New Mexico, 1919 Lomas Blvd NE, Albuquerque, NM 87131, USA}

\author[0000-0001-8812-0565]{Joseph E. Rodriguez}
\affiliation{Department of Physics and Astronomy, Michigan State University, East Lansing, MI 48824, USA}
\affiliation{Center for Astrophysics \textbar \ Harvard \& Smithsonian, 60 Garden St, Cambridge, MA 02138, USA}

\author[0000-0001-6213-8804]{Steven Villanueva, Jr.}
\affiliation{Department of Physics and Kavli Institute for Astrophysics and Space Research, Massachusetts Institute of Technology, Cambridge, MA 02139, USA}

\author[0000-0002-5951-8328]{Daniel J. Stevens}
\altaffiliation{Eberly Fellow}
\affiliation{Center for Exoplanets and Habitable Worlds, The Pennsylvania State University, 525 Davey Lab, University Park, PA 16802, USA}
\affiliation{Department of Astronomy \& Astrophysics, The Pennsylvania State University, 525 Davey Lab, University Park, PA 16802, USA}

\author[0000-0002-3481-9052]{Keivan G.\ Stassun} 
\affil{Vanderbilt University, Department of Physics \& Astronomy, 6301 Stevenson Center Lane, Nashville, TN 37235, USA}

\author[0000-0001-5160-4486]{David J. James}
\affiliation{ASTRAVEO LLC, PO Box 1668, MA 01931}

\date{\today}

\begin{abstract}
NASA's Transiting Exoplanet Survey Satellite (\tess) mission is expected to discover hundreds of planets via single transits first identified in their light curves. Determining the orbital period of these single transit candidates typically requires a significant amount of follow-up work to observe a second transit or measure a radial velocity orbit. In \citet{Yao:2019}, we developed simulations that demonstrated the ability to use archival photometric data in combination with \tess to ``precover" the orbital period for these candidates with a precision of several minutes, assuming circular orbits. In this work, we incorporate updated models for \tess single transits, allowing for eccentric orbits, along with an updated methodology to improve the reliability of the results. Additionally, we explore how radial velocity (RV) observations can be used to follow up single transit events, using strategies distinct from those employed when the orbital period is known. We find that the use of an estimated period based on a circular orbit to schedule reconnaissance RV observations  can efficiently distinguish eclipsing binaries from planets.  For candidates that pass reconnaissance RV observations, we simulate RV monitoring campaigns that enable one to obtain an approximate orbital solution. We find this method can regularly determine the orbital periods for planets more massive than $0.5 M_{\rm J}$ with orbital periods as long as 100 days.
\end{abstract}


\section{Introduction}
\label{sec:intro}

Follow-up observations of transiting exoplanets generally require precise ephemerides.  Any attempt to conduct transmission or emission spectroscopy, whether from the ground or from space, requires the ability to accurately predict future transits so as not to waste valuable telescope time or miss part of the event.  When candidate transiting planets have large uncertainties in their ephemerides, it becomes difficult to schedule and thus successfully perform follow-up observations. That is most often the case for long-period transit candidates, which we define here as candidates with orbital periods comparable to or greater than the time baseline of their originating transit survey. This situation was vividly demonstrated by \citet{Benneke:2017}, who found that the initially calculated ephemeris of the planet K2-18b ($P=33$ days with two transits shown in K2 data) was off by 2 hours before they recovered the correct ephemeris using the {\it Spitzer} telescope. More recently, \citet{Ikwut-Ukwa:2020} combined observations from Kepler and TESS for known K2 planets to reduce the uncertainty on future transit times from hours to minutes through 2030. Ephemeris errors of order hours would preclude the possibility of future observations from facilities such as the James Webb Space Telescope (JWST). 

Reliable ephemerides are also crucial to further dynamical studies of transiting planets.  \citet{Wang:2015} found that half of the ten long period exoplanets (periods between 430 days and 670 days) discovered by \kepler show transit timing variations (TTVs) ranging from $\sim$2 to 40 hours.  For planet discoveries in which the initial ephemerides are poorly constrained, additional transit observations might be needed to fix the ephemerides to permit later TTV analysis \citep{Dalba:2016,Dalba:2019}.  Obtaining such ephemerides from archival data saves valuable observing time, which can be helpful even in the cases of shorter-period planets that \tess is likely to detect.

The Transiting Exoplanet Survey Satellite (\tess) is designed to detect transiting planets orbiting bright stars across the whole sky. Launched in April 2018, \tess has observed 26 sectors and has discovered $\sim$ 2000 planet candidates so far. Among those, 51 have been confirmed and published. However, 90\% of the \tess planet discoveries have orbital periods shorter than 20 days\footnote{https://exoplanetarchive.ipac.caltech.edu/} due to the short observing time for most of the sky ($\sim27$ days). \citet{Cooke:2018} and \citet{Villanueva:2019} estimated that \tess will detect hundreds of planets with long orbital periods via single transits in their \tess light curves.

In \citet{Yao:2019} (henceforth Y19), we investigated the ability to recover the ephemerides of \tess single transit candidates using archival data from the Kilodegree Extremely Little Telescope (KELT) ground-based transit survey \citep{Pepper:2007, Pepper:2012}.  The process of using archival data to detect a signal originally revealed in later observations is sometimes called ``precovery".  In that work, we inserted simulated transit signals into KELT light curves, and explored the recoverability of the signals when combined with the information from \tess observations.  We found that a significant subset of large planets in long orbits that show single transits in \tess could be detected in KELT light curves, enabling precise measurements of their ephemerides.  This type of approach was successfully carried out by \citet{Gill:2020a} to recover the ephemeris of a single-eclipse \tess eclipsing binary using archival photometry from the WASP survey \citep{Pollacco:2006}.

There are many ways to consider how to best follow up and confirm single-transit candidates.  These approaches are different from those used to follow up transit candidates where the orbital period is known, since in those cases, photometric follow-up can be scheduled at specific times to catch future transits, and radial velocity (RV) observations can be timed to properly sample the orbital phase. In one approach to follow up single transits, \citet{Cooke:2020} explored the use of photometric versus spectroscopic follow-up observations to confirm single transits.  That analysis considered the use of three specific instruments: photometric observations using the Next Generation Transit survey (NGTS; \citealp{Wheatley:2018}), and radial velocity observations using the HARPS \citep{Mayor:2003} and CORALIE \citep{Queloz:2000} spectrographs.  To compare cases, \citet{Cooke:2020} considered the observing time required for a given instrument to achieve a detection of the planet past a given signal-to-noise ratio (SNR) threshold.  In \S \ref{sec:disc} below, we compare our approach to that work.  

In this paper, we improve upon the simulations in Y19 by incorporating more realistic transit models and orbital configurations. We also explore the use of RV observations of single-transit candidates to both eliminate certain types of false positives, and to confirm their planetary nature by measuring their orbits.  We do not compare single-transit follow-up strategies to those used for multi-transit candidates, since the unavailability of a known period means that the efficiency or expense of the efforts cannot be directly compared.  The paper is structured as follows: \S \ref{sec:precovery} updates the earlier analysis by deriving recovery rates of TESS single transit candidates with KELT photometry using realistic eccentricities for the simulated sample, and applying a more sophisticated calculation for the recovery rate. \S \ref{sec:EBs} discusses the use of RV observations to distinguish planetary systems from high-mass-ratio eclipsing binaries. \S \ref{sec:rv} presents simulations of RV observations to confirm planet candidates by constraining their orbital periods. We explore the results and implications of the result in \S \ref{sec:disc}, comparing this approach to other techniques, and in \S \ref{sec:sum} we summarize our findings.

\section{Updates to the Recovery Analysis}
\label{sec:precovery}

In Y19, we explored transit recovery of \tess single transits using KELT photometry to pre-cover the signals.  We simulated \tess single transits with known periods ranging from 13.5 to 300 days in circular and centrally transiting (i.e. equatorial) orbits, and the transit depth was assigned randomly from 3 mmag to 20 mmag in log space. Then we inserted the periodic transit signal into detrended KELT photometry, using a subset of 130,000 KELT light curves with RMS scatter below 30 mmag, out of all 5.8 million KELT light curves.  The selection of only low-noise KELT light curves effectively eliminates cases with stellar variability larger than the KELT photometric scatter.  We then used the Box-fitting Least Squares (BLS) algorithm \citep{Kovacs:2002} to try to recover the signal. We use the version of BLS with a fixed transit duration and a fixed $T_C$, since the parameters $T_C$ and duration for a given transit signal will be known with high fidelity from the TESS observations. The period was free to vary and we searched from 13.5 days to 300 days, with a frequency resolution of 300,000 (the number of trial frequencies scanned, evenly spaced in frequency). The recovered period was identified as the period corresponding to the strongest peak in the BLS periodogram. As the number of the light curves for the simulations is large, no visual inspection was conducted to check the recovered period. For a successful recovery of an inserted transit signal, we require that the percent difference between the recovered period and the inserted period be within 0.01\%. In that approach, we calculated the recovery rate for the KELT photometry for a range of transit durations and transit depths.  That is essentially an injection/recovery test of long-period transits with the KELT photometry under the assumption that the transit time and duration are known before searching for the signal in the KELT data.  Since planets with orbital periods longer than 10 days actually have a broad distribution of orbital eccentricities \citep{Marcy:2000, Winn:2015}, the duration and window coverage of transits is affected by the orbital eccentricity \citep{Barnes:2007, Burke:2008, Kane:2012, Kipping:2014}. Transiting planets with eccentric orbits will not have uniformly distributed arguments of periastron $\omega$; in eccentric orbits,  $\omega$ is more likely to be close to $\pi/2$ (as shown in Figure 4 from \citealp{Burke:2008}) and the planet is more likely to transit, which means the typical transit duration is shorter, for a fixed orbital period. Therefore, compared with circular orbits, planets in eccentric orbits will on average have fewer data points during the transit, which causes the SNR of the transit signals to decrease, and thus recovery rate for planets with more realistic (eccentric) orbits will be lower than cases with circular orbits. Therefore the recovery results from Y19 represent an overestimate of the recovery rates.

We address that issue by repeating the Y19 simulations but now consider eccentric orbits. To determine the orbital eccentricities in the updated simulations, we adopt the beta distribution with parameters $\alpha=0.867$ and $\beta=3.03$ from \citet{Kipping:2013}, and use the algorithm ECCSAMPLES \citep{Kipping:2014} to generate the orbital eccentricity and argument of periastron for each planet. Other parameter distributions such as orbital period, planetary radius, and orbital inclination are kept the same as in the circular orbital case.
Using the same criteria from Y19 to calculate the fraction of successfully recovered orbital periods (a period precision of better than 0.01\%), the average recovery rate declines as expected, and we find 5\% fewer recovered planets, where the overall recovery rate across all ranges of orbital period and planet radius drops from 33\% to 28\%.

We now explore ways to improve the utility of the recovery process.  We continue to operate under the assumption that the transit signal in the \tess light curve has a high enough signal-to-noise ratio for the transit duration and $T_{\rm C}$ to be calculated precisely, but without enough signal-to-noise ratio to strongly constrain the eccentricity just from the light curve.

Along with updating the Y19 analysis by incorporating eccentric orbits, we utilize the transit duration as observed by \tess in an additional way. In Y19 we required the recovered transit signal to have a similar duration as in the \tess light curve, but we did not constrain the range of possible periods to search based on the transit duration. Here, we consider the fact that the orbital period of a transiting planet can be estimated from the observed parameters of single transit assuming a circular orbit with a central transit \citep{Seager:2003, Yee:2008, Winn:2010}.  We do not assume that the orbits are necessarily circular, but rather make an educated guess that the orbit is not extremely eccentric, which allows us to more efficiently and reliably search a more limited range of possible periods, as described below. 

An alternate approach to our assumption of central transits is to use the transit shape as measured by \tess to estimate the impact parameter based on the transit shape in the \tess light curve.  However, that method relies on high SNR of the TESS detection, along with assumptions for the stellar limb-darkening.  Since we want to apply this method to as many of the \tess single transits as possible, we instead take the alternate approach of assuming central transits throughout this analysis.

Equation (7) in \citet{Yee:2008} expresses the relation between the orbital period and transit duration for a transiting planet. The way we incorporate this constraint is by calculating the orbital period from the transit duration under the assumption of circular and centrally transiting orbits ($P_{\rm cal}$) \citep{Winn:2010},
\begin{equation}
\label{eqn:Pcal}
P_{cal} = 365\;\mathrm{d}
\left(\frac{T_{dur}}{13\;\mathrm{hr}}\right)^{3}\left(\frac{\rho_{\star}}{\rho_{\odot}}\right).
\end{equation}
and requiring that the recovered orbital period from the BLS search ($P_{\rm BLS}$) be within the range of $P_{\rm cal}$ described below.  Even though $P_{\rm cal}$ is based on the assumption of a circular orbit, we can still use it to constrain the search for eccentric transits.  The transit duration in Eq. \ref{eqn:Pcal} is the full-width half-maximum of the transit in hours.

\begin{figure}[t]
\begin{center}
\makebox[\textwidth][c]{\includegraphics[width=1.0\textwidth]{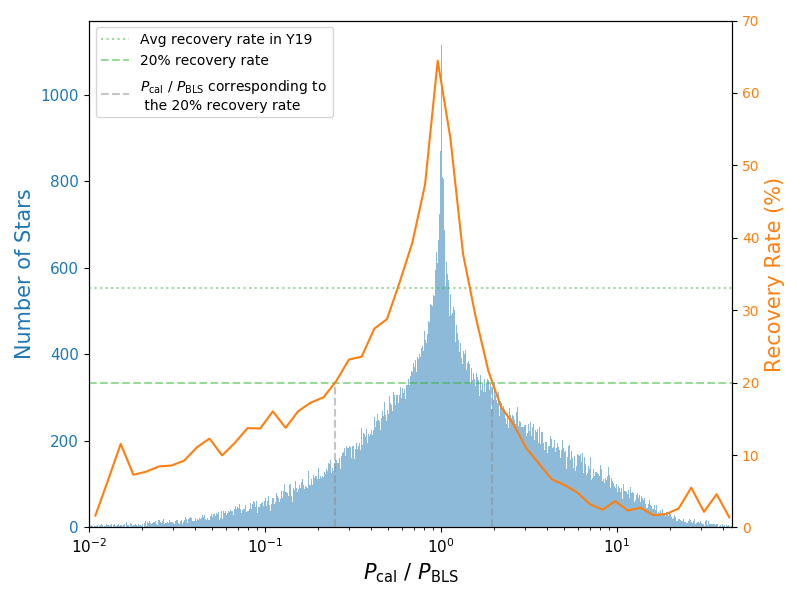}}
\caption{Distributions of the ratios between $P_{\rm cal}$ and $P_{\rm BLS}$ for eccentric orbits (blue) and the associated recovery rate as a function of the period ratio (orange). The green horizontal dotted line marks the average recovery rate in Y19 (restricted to circular orbits).  The green horizontal dashed line marks the 20\% recovery rate.  The vertical dashed grey lines indicate the period ratio corresponding to the 20\% recovery rate.}
\label{fig:hist}
\end{center}
\end{figure}

Figure \ref{fig:hist} shows the distribution of the ratio $P_{\rm cal}/P_{\rm BLS}$, along with the recovery rates. Since the BLS search is performed across period space for all light curves, we expect some spurious signals at all period ranges.  By requiring $P_{\rm BLS}$ to be sufficiently close to $P_{\rm cal}$, we can eliminate many spurious cases from the analysis and improve the reliability of the surviving signals.
We thus implemented an additional cut in the analysis by requiring that $P_{\rm cal}/P_{\rm BLS}$ to be between 0.25 and 1.96, which effectively removes cases where the expected recovery rate is below 20\%.  We selected that particular cutoff to reflect a subjective judgement regarding the balance between improving the overall recovery rate, and retaining a large fraction of the total sample.  We then calculated the overall recovery rate for the remaining simulated light curves.

Using this constraint, the overall recovery rate improves by $\sim$12\%, from 28\% to 40\%, as a result of excluding cases where the period calculated from BLS is likely inconsistent with the transit duration.  In this case, that restriction excluded $\sim$41\% of the original set of simulated light curves, finding the BLS-derived period to be unlikely to be consistent with the transit duration. This recovery rate is also higher than the overall recovery rate found in Y19, even though we now account for eccentric orbits.  The improvement in the recovery rate from 28\% to 40\% does not mean that we can detect more long-period planets from \tess, but that the reliability of the results from searching the KELT data is more trustworthy.

\begin{figure}[htp]
\makebox[\textwidth][c]{\includegraphics[width=0.95\textwidth]{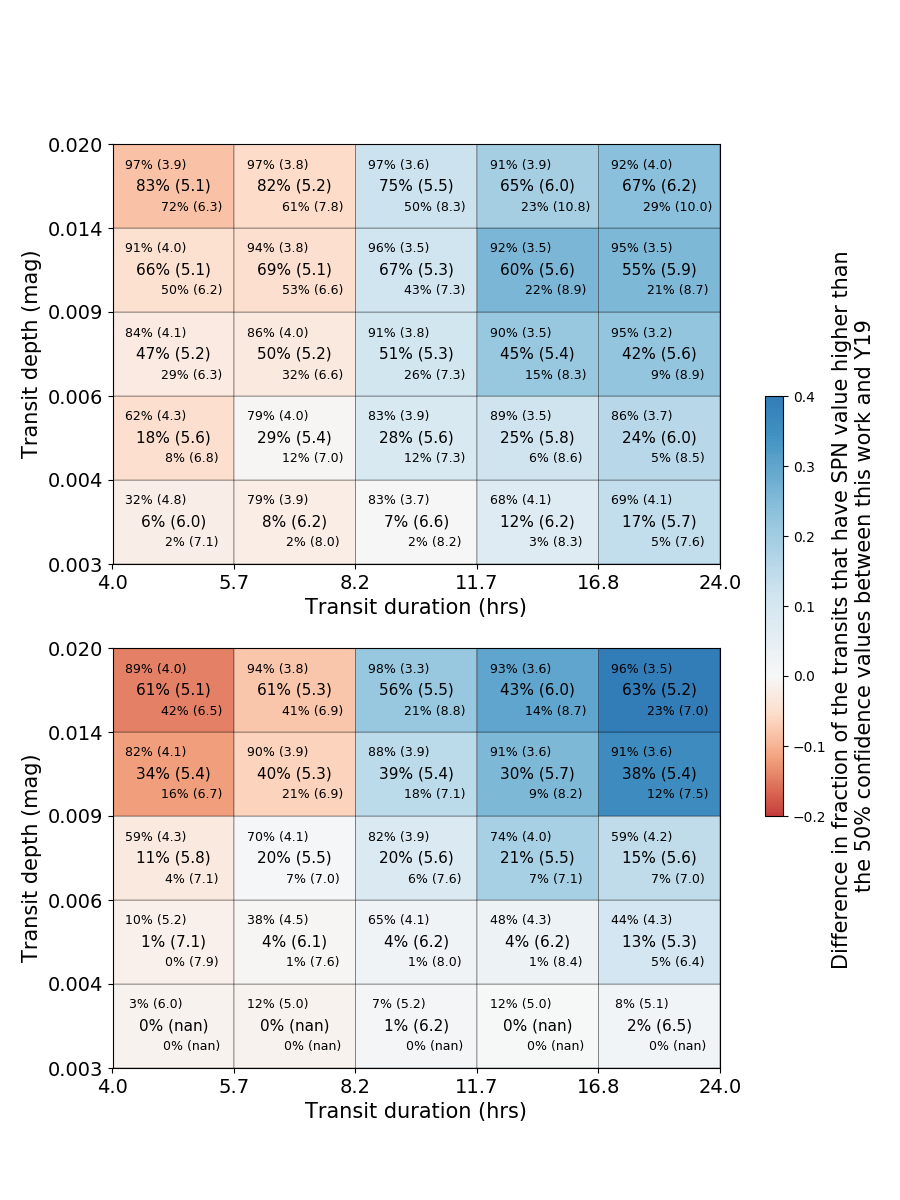}}
\caption{The fractions of KELT-North (top) and KELT-South (bottom) light curves that pass SPN thresholds in transit depth / transit duration bins, and the corresponding SPN values in parentheses. In each box are three percentage values and corresponding SPN values. The percentages reflect the fraction of KELT light curves in that bin of depth/duration space such that if the SPN value is greater than the indicated value, there is a given likelihood that the recovered period is correct. Those likelihoods are 10\%, 50\%, or 90\%, from upper left to lower right in each bin.  The color bar indicates the fractional change in the fraction of KELT light curves that pass the 50\% confidence threshold between this analysis and the results from Y19, (discussion in text). The items marked `nan' represent cases where none of the inserted signals were recovered in the KELT light curves at the corresponding confidence level for the depth and duration ranges for that bin.}
\label{fig:spn}
\end{figure}

To show the result in a more useful form for follow-up observers, we adopt the signal-to-pink noise ratio (SPN) as a signal strength criterion to estimate the confidence of the recovery result. The SPN ratio is a variant of the S/N computed using a “pink” total noise that includes both uncorrelated (``white") and correlated (``red") noise sources. More details of the definition of SPN can be found in \citet{Hartman:2016}, and details of how we utilize this statistic are in section 4.2 of Y19.  In short, we calculate the empirical SPN threshold where a given percentage of all test cases are successfully recovered, and use that as the "recovery confidence".  For instance, in a given range of transit depth and duration, if we find that 90\% of all test signals are recovered in the KELT light curves, then that SPN value is our 90\% confidence threshold.  Similarly to Figure 13 in Y19, Figure \ref{fig:spn} shows the SPN thresholds for 10\%, 50\% and 90\% recovery confidence with the fraction of KELT light curves that are above those thresholds across transit depth versus transit duration, using the same parameter ranges as in Y19.  In that figure, we show the results separately for the KELT-South and KELT-North data sets, since the KELT-North dataset generally includes more epochs spread over a longer time baseline as compared to the KELT-South dataset.  For example, consider the center bin in the upper panel.  The lower right row within that bin refers to the SPN required for a transit to be recovered 90\% of the time.  We find that SPN value to be 7.3, so that if a transit signal with an SPN value greater than 7.3 is seen, an observer can have 90\% confidence the signal is real.  The percentage in that row indicates that of the KELT light curves with inserted transit durations and depth in that corresponding range, 26\% yielded SPN values greater than 7.3.

The color of the cells in Figure \ref{fig:spn} indicate whether the updated analysis yields better (blue) or worse (red) recovery rates compared to Figure 13 in Y19. That is, the colors indicate the differential improvement (or decline) in recovery rates compared to the assumption of a circular orbit.  The figure conveys two sets of information - the absolute recovery rates, using the methodology described above, as well as the relative increase or decrease in recovery rates across parameters space compared to Y19.

The colors in the figure reflect the impact of the two key changes we have made.  The use of eccentric orbits decreases the recovery rate throughout the analysis.  That is because for the same transit duration, the real orbital periods for eccentric transiting planets are generally longer than the ones with circular orbits. Therefore the SNR of the transit signal in the light curves is lower (the converse of the effect mentioned above) when we are observing multiple transits.  On the other hand, the use of the additional information coming from the transit duration using $P_{\rm cal}$ improves the reliability of the recovery (and thus the overall recovery rates) due to constraining the period search range. Figure \ref{fig:spn} shows that for transit durations shorter than 8 hours, the former effect dominates, while at longer transit durations, the latter effect dominates.


\section{Distinguishing Eclipsing Binaries From Planets}
\label{sec:EBs}

Although in this analysis we have inserted transit-like signals into the KELT light curves as a phenomenological feature, agnostic as to the physical cause of the signal, we now explore not just how to detect those photometric signals, but also how to distinguish their physical cause. 

One feature of the mass-radius relationship for stars and planets is that the onset of electron degeneracy in the core leads to a flat mass-radius curve from the massive planet regime to the small star regime \citep{Zapolsky:1969, Chabrier:2000, Lynden-Bell:2001, Fortney:2007}. That is, compact objects with masses between Jovian planets and late-type stars all have radii of roughly 1 $R_{\rm J}$, despite an almost two orders of magnitude range in masses. Among our simulation samples, $\sim$65\% of planetary candidates have radii larger than $1 R_{\rm J}$.
That means a large number of transit signals in our simulations may be caused by eclipsing binaries instead of planets.  They can include a larger star being eclipsed by a later-type dwarf star, yielding a primary eclipse depth comparable to that of a transiting giant planet, or an eclipsing binary with a large primary transit that is diluted by blending with a nearby bright star. In this section we will concentrate on ways to identify the first of these scenarios.  We will ignore transit-like signals  caused by blended eclipsing binaries.  Those types of false-positive scenarios can be addressed through careful analysis of the TESS pixel data by looking for centroid shifts (\citealp{Bryson:2013}, Guerrero, et al.\ submitted), or via photometric monitoring with higher angular resolution or other spectroscopic techniques (see \citet{Collins:2018} for a discussion on vetting different false positive types).

One tool for distinguishing transiting planets from eclipsing binaries is the use of radial velocity (RV) observations. Here we explore how to conduct RV observations in the case where the ephemeris of the transit signal is unknown or poorly constrained, as in the case of single transits.  This process often involves two stages of follow-up observations.  The first stage, often referred to as reconnaissance spectroscopy, involves two spectroscopic observations.  These observations are taken at the predicted orbit's quadrature phases, and can identify SB1s, and visual inspection of the spectra can identify SB2s.  Once those scenarios are ruled out, further RV observations are obtained over the full phase of the orbit to characterize the orbital elements, e.g., the RV semiamplitude, argument of periastron, and eccentricity. 

In the case of an unknown orbital period, a single spectroscopic observation can still identify SB2-type cases, but without knowing the predicted quadrature times, it is not obvious how to perform reconnaissance spectroscopy so as to rule out EBs. However, if the system has a large semi-amplitude, two random observations could show a large offset indicating the presence of an EB. There has been extensive work conducting both RV and photometric follow-up observations of single-transit candidates from the NGTS team, recovering the ephemerides of two EBs that showed single transits in \tess photometry \citep{Lendl:2020,Gill:2020c}, along with the discovery of a long-period planet from what was initially thought to be a single transit in \tess data, which was later revealed to contain an additional transit initially obscured by scattered light in the photometry \citep{Dalba:2020, Gill:2020b}.  Here we investigate a comprehensive approach to RV follow-up.  

The dataset used in Y19 and in \S \ref{sec:precovery} consists of KELT light curves in which we inserted a set of simulated transit signals.  
In this section, we do not use any light curves, but we apply the properties of the simulated planetary systems from that analysis, namely the orbital periods, radii of eclipsing objects, eccentricities, and argument of periastron, and the masses and radii of the associated stars, with the stellar parameters taken from the TESS Input Catalog (TIC-8).

We first introduce EB cases to the simulated signals.  Note that the signals were drawn from a range of empirical properties (depth, duration, period, etc.), without reference to physical properties (companion mass, radius, semimajor axis, etc.).  We selected $\sim$48,000 targets from the KELT-North sample of stars described in \S \ref{sec:precovery} that have radii of the transiting body that are larger than $1 R_{\rm J}$, as inferred from the simulated transit depths and the estimated stellar radii from the TIC-8.  This subset of the full light curve simulation includes the candidates for which there is ambiguity about the physical nature of the eclipsing body.  
In the process of assigning masses, we now consider two cases: if the transiting objects are all planets, the mass ($M_p$) was randomly assigned between $0.2 M_{\rm J}$ and $3M_{\rm J}$ in log space; and if the transiting objects are stars, the mass ($\rm M_{\sec}$) was calculated based on the mass-radius relation in the stellar region derived from \citet{Chen:2017}. We are essentially assuming that objects with masses between $3 M_{\rm J}$ and $80 M_{\rm J}$ are rarer than planets or stars, i.e. that there exists a ``Brown-dwarf desert" \citep[e.g.,][]{Grether:2006}.  The semi-amplitude ($K$) of the radial velocity curve for planets was calculated using the above parameters:
\begin{equation}
\label{eqn:K}
K = 28.4 \;\mathrm{m/s}
\left(\frac{P}{1\;\mathrm{yr}}\right)^{-1/3}\left(\frac{M_{\rm p}}{M_{\rm J}}\right)\left(\frac{M_{\star}}{\rm M_{\odot}}\right)^{-2/3}\left({1-e^2}\right)^{-1/2}.
\end{equation}
For the case of binary stars, we replace $M_{\rm p}$ with $M_{\sec}$ and $M_{\star}$ with $(M_{\star}+M_{\sec})$ in Equation \ref{eqn:K}. Note that, since we know these systems are seen nearly edge-on because they exhibit transits or eclipses, Equation \ref{eqn:K} assumes $\sin{i}$ $\simeq$ 1.
We therefore ignore the $\sin{i}$ dependence on $K$ in Equation \ref{eqn:K}. 
We used the python package RadVel \citep{Fulton:2018} to synthesize RV curves based on those parameters. 

While the observational strategies of RV follow-up can vary significantly, we simulated a plausible follow-up approach that might be undertaken by a mid-size RV survey facility such as TRES \citep{Szentgyorgyi:2007}, MINERVA \citep{Swift:2005}, MINERVA-Australis \citep{Addison:2019}, the APF \citep{Vogt:2014}, or a similar facility. We assume that the time interval between the TESS observations of the single transit and the first RV observation is three months, which is consistent with the TESS data release procedure, and the time required to identify candidates and begin scheduling RV observations.  We also tested lengths of time intervals of 1 month and 12 months, respectively, and found very little effect on the results.  

The RV precision of an observation depends on the telescope, the instrument and various stellar properties such as brightness, effective temperature, and rotational velocity, among other factors.  Although we explored the possibility of simulating the RV precision for each observation based on the stellar properties and assumed noise model for a given instrument, we ultimately found such an effort to be unfeasible.  Even for a specific telescope and noise model, there is a spread in RV precision for stars of a given brightness, with many ways that the stellar properties can influence the RVs, such as the effects of stellar rotation, spot coverage, and chromospheric activity. Therefore, we decided to adopt a uniform RV precision of 20 m/s in the simulation, assuming the stellar rotation velocities are lower than 10 km/s. These assumptions were based on the statistics of real observational data from the CHIRON spectrograph \citep{Tokovinin:2013}.

Since the true orbital period is not precisely known from TESS single transit light curves particularly when assuming eccentric orbits, we modeled the approach of an observer making an educated guess about the quadrature times based on $P_{\rm cal}$.  We then determined the RV values at those times according to the model RV, and then added offsets to the calculated RV using values drawn from a Gaussian distribution with a width of 20 m/s. We then approximate the RV semi-amplitude as half the difference in the simulated RV measurements at the estimated times of quadrature.  This makes the inherent assumption that the orbit is approximately circular, and thus the quadrature times are approximately known, and that the semiamplitude is just one half the difference between the measurements at quadrature (ignoring measurement errors). 

We do not consider whether the calculated quadrature time is during local night, but since the potential orbital periods are known to be much longer than 24 hours, we assume that observations will take place within 12 hours of the calculated quadrature, and that issue does not have a major impact on these results.  Another potential issue with this approach is that we assume that all planet candidates included are observable at some point in the night through the campaign.  If targets are randomly distributed in right ascension, then some targets will inevitably be unobservable for significant lengths of time.  But with a significant number of total candidates, the observers can select a subset that should be observable through the whole campaign.

\begin{figure}[t]
\begin{center}
\makebox[\textwidth][c]{\includegraphics[width=1.0\textwidth]{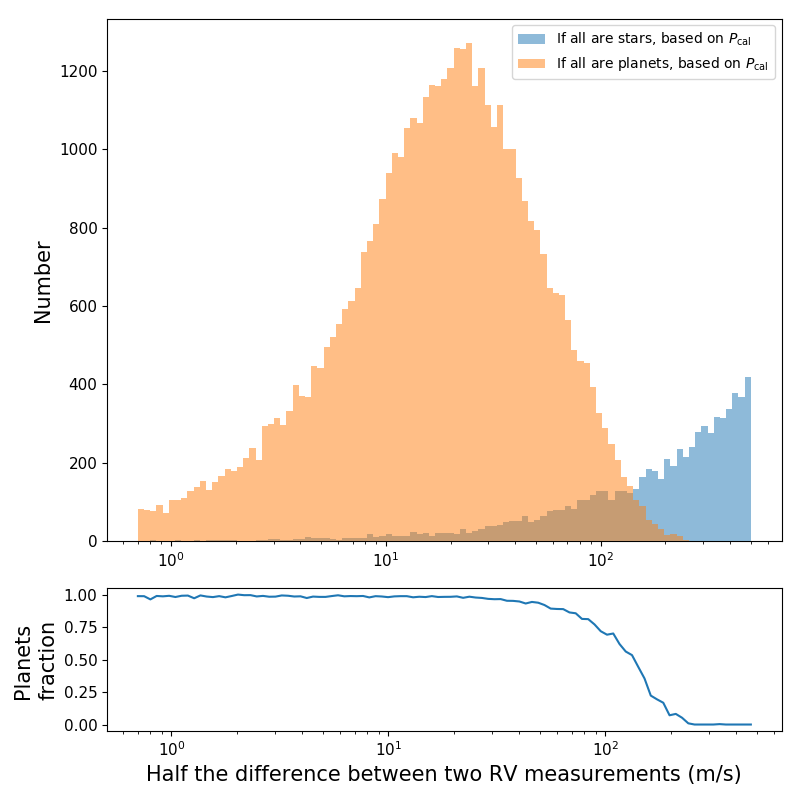}}
\caption{The upper panel shows the distributions of half the difference between two RV measurements at estimated quadrature times for the case of planetary companions (orange) and a portion of the distribution for stellar companions (blue).  The difference between the RV observations is a rough proxy for the semiamplitude.  The lower panel indicates the fraction of the samples at a given calculated semi-amplitude in which the candidates are planets.}
\label{fig:EBs}
\end{center}
\end{figure}

In the upper panel of Figure \ref{fig:EBs}, we show the distribution of estimated $\rm K$ for the planet and star samples, respectively. The lower panel indicates the fraction of the samples at a given estimated $\rm K$ in which the candidates are planets. If the difference between the two measured RVs is smaller than 50 m/s, the transit signal has a less than 10\% probability of being a stellar companion, and more RV follow-up is merited to determine the full dynamical orbit. If the RV difference is larger than 200 m/s, it is likely to be an EB with greater than 90\% probability.  For RV differences between those ranges, additional observations are required to identify potential false positives.

\section{Radial Velocity Detection of Planets}
\label{sec:rv}

RV observations to dynamically detect planets are generally conducted in one of two ways. If the orbital period of a system is known from multiple transits, RV observations can be obtained across a full orbital phase, to measure the orbital reflex motion of the star due to the planet \citep[e.g.][]{Burt:2018, Medina:2018}. If the orbital period is not known, as in a blind RV search, one star or a set of stars can be monitored by (semi)regular RV observations until orbital motion is seen in a phase-folded search of the RV data. 
In these situations, RV monitoring, which can detect the gravitational reflex motion that a planet induces on its host star throughout the entire orbit, can be used to determine the orbital parameters.

In this section we conduct a simulation of what such an RV campaign might look like, and we examine the distribution of stellar and planet parameters for the systems for which that campaign could successfully detect the orbit.  It should be noted that RV confirmation of a single-transit candidate must operate quite differently than for a multiple-transit candidate.  Since the photometry provides no direct measure of the orbital period, the RV observations must effectively operate as a blind RV search, as in the second case noted above.  The approach described here aims to provide a more efficient way to conduct that sort of search.

To create the simulation, we randomly selected 10\% of the stars from the sample in \S \ref{sec:precovery} that have $T_{\rm eff}$ between 4000K and 7400K (i.e., FGK type stars) which are best suited for RV follow-up. This results in a sample of $\sim$5600 stars in the KELT-North set and $\sim$5400 stars in the KELT-South set. Note again that the data set involved in this analysis consists of the simulated transit signals, not the KELT light curves themselves.  We start with the empirical signal properties, and translate those into associated physical properties of the transiting companions, as we did in \S \ref{sec:EBs}.  In this case, if the transiting objects have radii larger than $1 R_{\rm J}$, the mass was randomly assigned between $0.2 M_{\rm J}$ and $3 M_{\rm J}$ in log space; if planets have radii smaller than $1 R_{\rm J}$, the mass was calculated based on the mass-radius relation in the Neptunian region derived from \citet{Chen:2017}.

The RV observations are simulated over a 3-month time span.  That duration was chosen as a plausible length of time for a dedicated RV campaign of this type.  A natural consequence of this is that systems with true periods longer than 3 months will be incompletely sampled, which we discuss below.  The parameters $P$, $e$, $\omega$ and $T_{\rm C}$ used to generate the RV curves are the same as those from \S \ref{sec:precovery}, in which we have included eccentric orbits, and we again assume an RV precision of 20 m/s. We randomly selected 7 days each month in which one observation of the target star is obtained each night, for a total of 21 observations per star over that time frame.  We assume that the stipulation of 7 RV observations per month can be met even in the presence of weather, which we account for by the random placement of the observations within each month. We assigned RV values according to the theoretical RV curve, and then added Gaussian noise with a distribution width of 20 m/s to the RV points.  

It should be noted that this approach does not account for dynamic scheduling of upcoming RV observations during a campaign based on an evaluation of the RV results up to that point.  That topic has been explored extensively before \citep[e.g.][]{Kane:2007,Ford:2008,Loredo:2011}, but not in the specific case of a singly-transiting planet where a date of conjunction and the planet size are known, but not the orbital period.  In a related analysis, \citet{Cabona:2020} examined the efficacy of different scheduling strategies when conducting follow-up RV observations of small TESS targets using the ESPRESSO instrument \citep{Pepe:2014} on the VLT.  However, that work considers only transit candidates with known periods, and so addresses a different scientific question than considered here.

There are various software tools that perform RV fitting, such as RadVel \citep{Fulton:2018} and the Joker \citep{Price:2017} and EXOFASTv2 \citep{Eastman:2019}. 
In the next step, we conducted a maximum likelihood Keplerian model fit as implemented in the RadVel package for the synthesized RV data for each target. In the fitting process, the transit time $T_{\rm C}$ was fixed to the simulated value as would be measured from the TESS single transit, and we set boundaries from 13.5 days to 300 days for the orbital period. Other parameters ($K$, $e \sin \omega$, $e \cos \omega$) were free to vary. For the initial guess of the orbital period, we used $P_{\rm L-S}$ from a Lomb-Scargle period search of the simulated RV data. The initial guess of $K$ was calculated based on the assumption of a circular and centrally transiting orbit with $P_{\rm L-S}$, known stellar mass $M_{\star}$, and $M_{\rm p}$ calculated based on the $M_p-R_p$ relation from \citet{Chen:2017} as described above. 

There are a number of ways to consider the precision on the period of an exoplanet derived from RV observations.  For the purposes of confirming a planet, a fractional period precision of tens of percent may be sufficient.  For conducting intensive transit follow-up however, such as atmospheric characterization, an absolute precision of better than 30 minutes would typically be required.  For the analysis here, we consider an RV-fitted period to be correct if the fractional difference between the RV-fitted period $P_{\rm RV,fit}$ and $P_{\rm real}$ is smaller than 5\%.  While that precision would not be sufficient for atmospheric observations, or even for photometric ephemeris confirmation in some cases, it would be more than sufficient to identify good candidates for long-period transiting planets from the TESS sample of single-transit candidates.  At that point, an observer can conduct additional RV observations beyond the campaign envisioned here, with higher cadence, or with another facility with greater RV precision to bring the measured period from the new RV observations to a level sufficient for photometric ephemeris confirmation to obtain the higher absolute precision on the transit time needed for intensive transit observations.  For similar reasons, we consider an RV-fitted $K$ to be correct if the fractional difference between $K_{\rm fit}$ and $K_{\rm real}$ is smaller than 50\%. Our goals at this stage are to obtain approximate orbital solutions and to differentiate likely planets from false positives, rather than obtaining a complete system solution at the end of the campaign.

It is the case that while the simulated RV campaign lasts for only 90 days, we search for orbital periods out to 300 days.  We made this choice for two reasons.  First, it is possible to obtain an RV detection of the orbital signal with only partial phase coverage, although the fractional period precision of the resulting fit is typically quite poor.  Since a real campaign, regardless of its duration, will end up observing some systems with long orbital periods, we wanted to test the ability to extract such signals.  In most cases of periods much longer than the campaign duration, only a linear trend will be identified.  We discuss such cases in more detail below.

\begin{figure}[t]
\begin{center}
\makebox[\textwidth][c]{\includegraphics[width=1.0\textwidth]{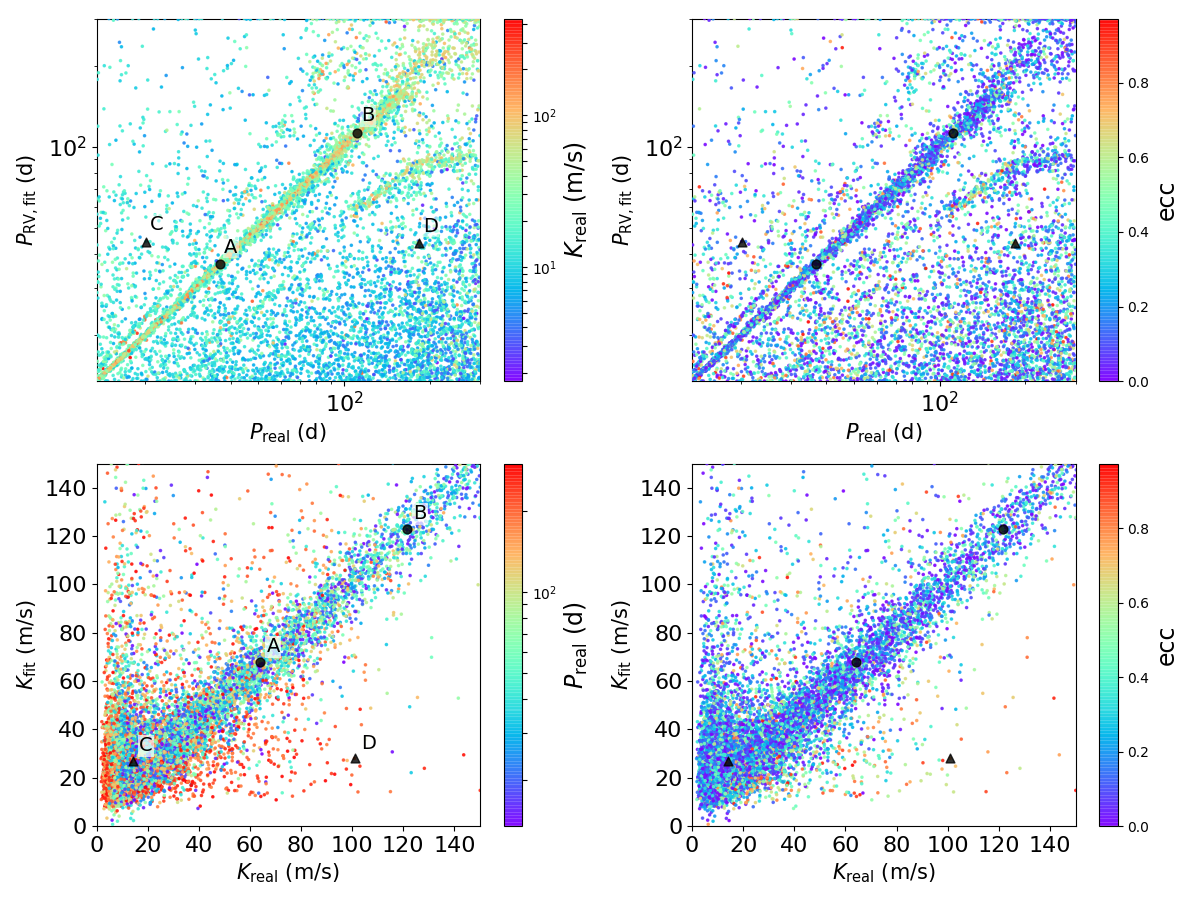}}
\caption{Scatter plots for the simulated samples, comparing the real orbital periods and semi-amplitudes with their fitted values, colored by different real parameter values.  We indicate the real and fitted values for the 4 examples in Figure \ref{fig:examples} with filled black points.  In the upper plots, the populations that run along side the diagonal represent cases where the fitted period is off by a factor of 2 or 1/2.}
\label{fig:pke}
\end{center}
\end{figure}

Figure \ref{fig:pke} shows the distributions of the real and fitted parameters $P_{\rm real}$, $P_{\rm RV,fit}$, $K_{\rm real}$, $K_{\rm fit}$ and $e_{\rm real}$.  There is clearly a significant population along the diagonal in each panel which represents the correctly fitted samples, along with clusters running along the diagonals in the period plots which indicate cases off by factors of 2 or 1/2. As expected, systems with high eccentricity and small semi-amplitudes tend to be recovered least well.  Systems with long orbital periods are also poorly recovered, although that is partly due to the fact that the 90-day span of the RV observations is shorter than the orbital period for some of the systems. By comparing the fitted parameters with real values, we found the median absolute deviation (MAD) of ($P_{\rm RV,fit}$ versus $P_{\rm real}$) is 7.2 days and the MAD of ($K_{\rm fit}$ versus $K_{\rm real}$) is 8.2 m/s. Although these intermediate results provide a rough estimate of the ephemerides of the candidates, these ephemerides are not sufficiently precise to schedule follow up observations during transits. We improve upon these results below by identifying the more reliable fitted periods.

\begin{figure}[t]
\begin{center}
\makebox[\textwidth][c]{\includegraphics[width=1.0\textwidth]{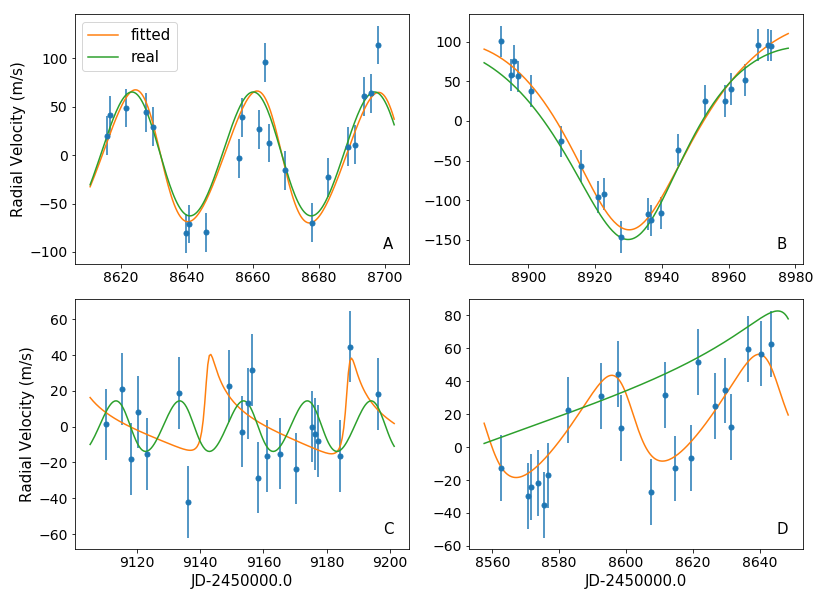}}
\caption{Two successfully fitted RV samples (top panel) and two failed fitted samples (bottom panel). All these four examples are identified in Figure \ref{fig:pke} and Figure \ref{fig:fap}.}
\label{fig:examples}
\end{center}
\end{figure}

Figure \ref{fig:examples} shows two examples that were successfully fitted and two examples where the fitting failed.  We indicate the locations of these four examples on the plots in Figure \ref{fig:pke}.  The two successful fits (cases A and B) show good agreement between the true periods and the fitted periods for both shorter and longer orbital periods.  Example C shows a case where the true semiamplitude is smaller than the scatter in the RV points, and example D shows a case where the true period is significantly longer than the time baseline of the RV data.

\begin{figure}[t]
\begin{center}
\makebox[\textwidth][c]{\includegraphics[width=1.0\textwidth]{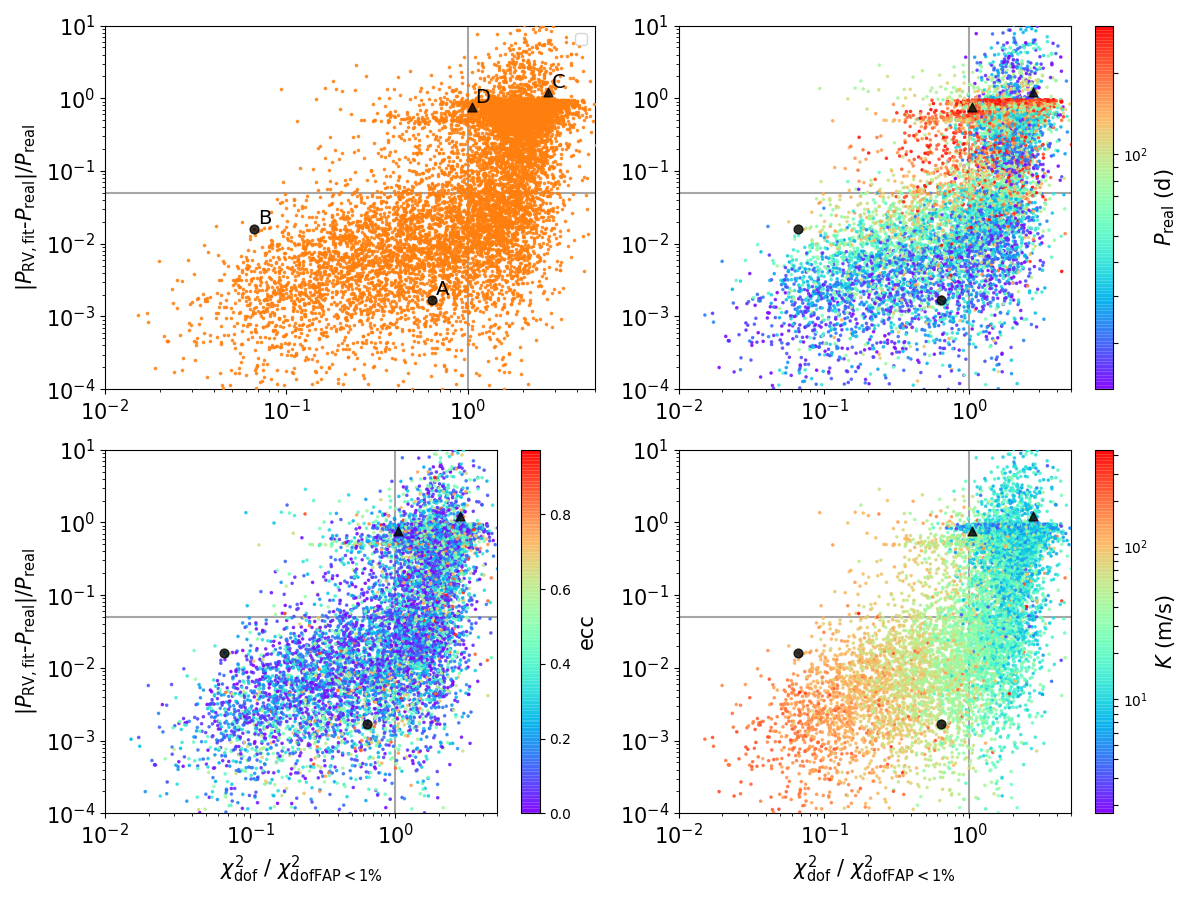}}
\caption{Scatter plots of the percentage difference between the fitted period $P_{\rm RV,fit}$ and $P_{\rm real}$ versus the 1\% FAP ratio $\chi_{\rm dof}^{2}$/$\chi_{\rm dof{FAP<1\%}}^{2}$. The color bars represent the real value of period, eccentricity, and semi-amplitude.}
\label{fig:fap}
\end{center}
\end{figure}

We compute the reduced chi square ($\chi_{\rm dof}^{2}$) between the best-fit model and the simulated data. By inspecting the fitting results, we found nearly all samples have $\chi_{\rm dof}^{2}$\textless2, which indicates that $\chi_{\rm dof}^{2}$ is not an efficient tool to measure the confidence of the fitted RV period.  That situation arises because the amplitudes of the RV signals in our sample are often not that much larger than the per-point errors that we are modeling.  As $\chi_{\rm dof}^{2}$ is not an efficient tool to measure the confidence of the fitted RV period, we instead conduct a False Alarm Probability (FAP) test of the RV fits. For each star in the simulation, we ran 100 iterations of the above analysis after randomizing the RV observations and the observing times. We selected the smallest reduced chi square $\chi_{\rm dof{FAP<1\%}}^{2}$ to indicate the 1\% FAP value for that star. Figure \ref{fig:fap} shows the distribution of period recovery accuracy compared to the ratio between $\chi_{\rm dof}^{2}$ and $\chi_{\rm dof{FAP<1\%}}^{2}$, which we refer to as the 1\% FAP ratio. The plots in that figure include horizontal lines indicating the 5\% accuracy threshold on period recovery, and vertical lines at the FAP ratio of unity indicating whether the $\chi_{\rm dof}^2$ fit was larger or smaller than the 1\% FAP threshold.  The region to the right of the FAP threshold indicates systems where the RV fit is not very reliable, and tends to include long-period and low-amplitude systems.  The region to the left of the FAP threshold indicates reliable results, and we find that 85\% of those results yield orbital periods that are within 5\% of the true period.  Of the 15\% cases that are considered reliable from the FAP cut but do not match the orbital period (the upper left region in the first panel), 82\% of those have true periods longer than the observing campaign, indicating that we see a broad trend without the leverage to accurately measure the period.  We conclude that the 1\% FAP ratio can be used as an efficient tool to measure the reliability of the fit. Similar to Figure \ref{fig:pke}, systems with a relatively short period
and large semi-amplitude are most likely to be correctly fitted, as expected.

\begin{figure}[t]
\begin{center}
\makebox[\textwidth][c]{\includegraphics[width=1.0\textwidth]{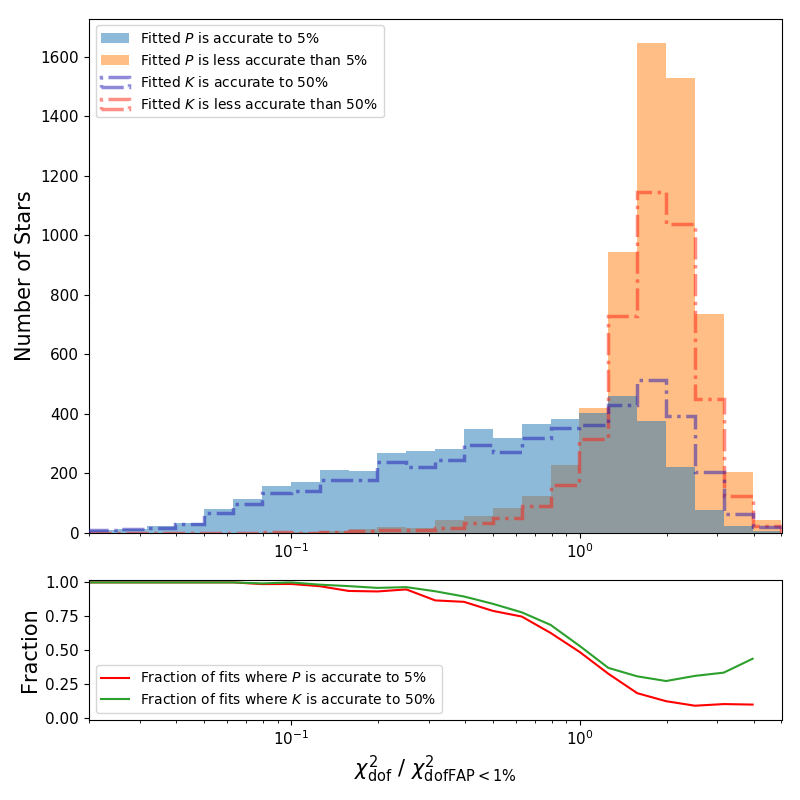}}
\caption{Distribution of the 1\% FAP ratio ($\chi_{\rm dof}^{2}$/$\chi_{\rm dof{FAP<1\%}}^{2}$) for tested RV fits, colored by whether the fitted period was close to the correct period based on a period precision of 5\%, with the grey color indicating the overlap of the two. The dot-dashed histogram represents the cases where the semi-amplitude was near the real value. The lower panel indicates the fraction of the tested RV fittings at a given bin in 1\% FAP ratio in which the period and semi-amplitude were correctly fitted.  At a 1\% FAP ratio of about 1, half of the test samples have well-fitted periods and semi-amplitudes.}
\label{fig:FAP_hist}
\end{center}
\end{figure}

Figure \ref{fig:FAP_hist} provides a different way to view the utility of the 1\% FAP ratio.  The figure shows the distribution of the 1\% FAP ratios for all test cases, with one set of values for the period fit, and another for the semi-amplitude fit, separated according to passing the fit threshold.  At a 1\% FAP ratio around unity, half of the cases were correctly fitted, and at a 1\% FAP ratio around 0.3, nearly all samples are correctly fit. These distributions provide confidence metrics for the fitted period and semi-amplitude. 

\begin{figure}[t]
\begin{center}
\makebox[\textwidth][c]{\includegraphics[width=1.0\textwidth]{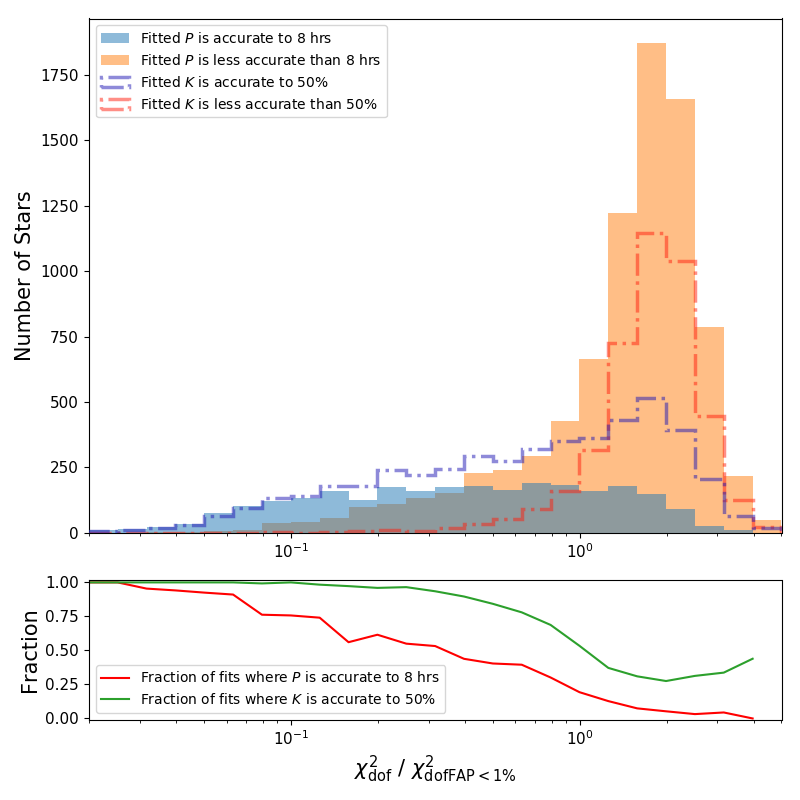}}
\caption{Distribution of the 1\% FAP ratio ($\chi_{\rm dof}^{2}$/$\chi_{\rm dof{FAP<1\%}}^{2}$) for tested RV fits, colored by whether the fitted period was close to the correct period based on an 8-hour absolute accuracy. The dot-dashed histogram represents the cases where the semi-amplitude was within 50\% of the correct value.}
\label{fig:FAP_hours}
\end{center}
\end{figure}

As discussed above, the precision of a fitted period has different implications depending on the goals of an investigation.  While a given fitted period might have a low fractional uncertainty, if an observer wants to establish the true transit ephemeris, they will often want to obtain a fitted period with a specific absolute uncertainty, corresponding to the time span over which the second transit is expected at a confidence of, say, 1$\sigma$ or 68\%, so as to schedule follow-up photometry to measure the transit. Figure \ref{fig:FAP_hours} shows another version of Figure \ref{fig:FAP_hist}, by using a fixed absolute uncertainty on the period of 8 hours rather than a fixed fractional uncertainty on the period.  This version provides a more useful reference for observers to plan photometric follow-up observations, assuming an 8-hr night for observing.  Specifically, if an observer wants to obtain follow-up photometry for targets with an RV-fitted period with precision smaller than 8 hours, with at least 50\% confidence, then they should select targets with a 1\% FAP ratio smaller than 0.3 (see \S \ref{sec:disc} below for details on implementing this process.)

\begin{figure}[t]
\begin{center}
\makebox[\textwidth][c]{\includegraphics[width=1.0\textwidth]{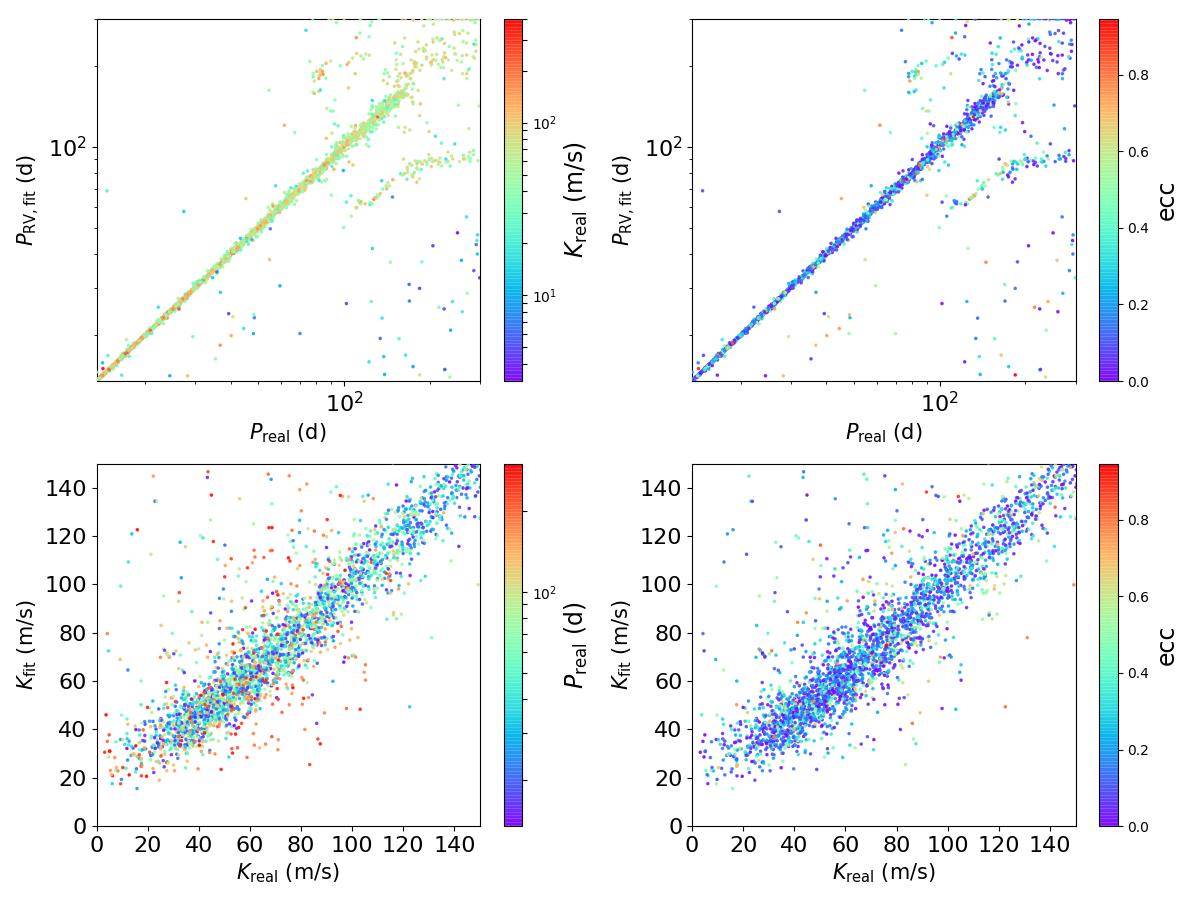}}
\caption{Scatter plots for the simulated samples, similar to Figure \ref{fig:pke}.  In this figure, only cases where the 1\% FAP ratio is smaller than 1 are displayed.}
\label{fig:pke_goodfap}
\end{center}
\end{figure}

\begin{figure}[t]
\begin{center}
\makebox[\textwidth][c]{\includegraphics[width=1.0\textwidth]{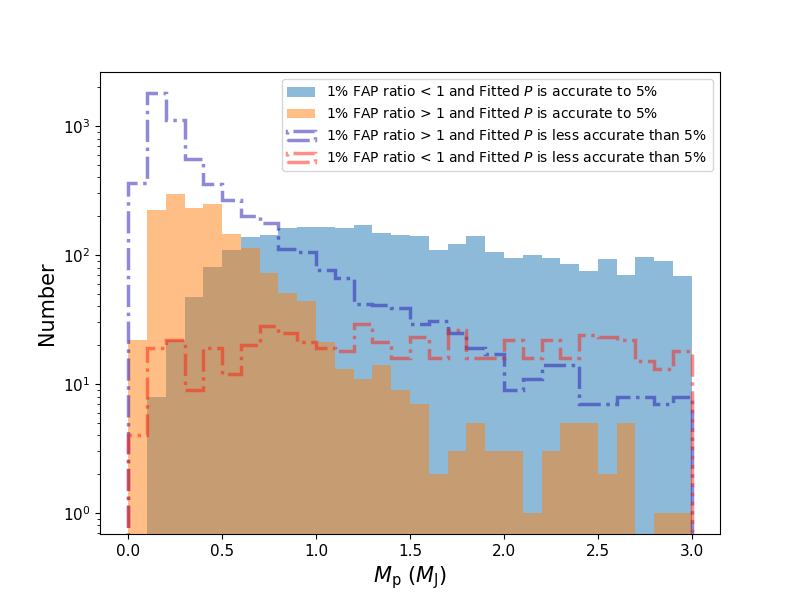}}
\caption{Distribution of the planet masses for the successfully and unsuccessfully recovered simulated systems.}
\label{fig:Mp}
\end{center}
\end{figure}

Figure \ref{fig:pke_goodfap} reproduces Figure \ref{fig:pke}, showing the distributions of the real and fitted parameters, but this time only for the samples with 1\% FAP ratio smaller than 1. In this case, the median absolute deviation (MAD) of  the difference between $P_{\rm RV,fit}$ and $P_{\rm real}$ is 0.3 days, and the MAD of the offset between $K_{\rm fit}$ and $K_{\rm real}$ is 6.5 m/s. There remain some clusters of fits that are off by a factor of 2 or 1/2. This result shows the clear improvement in the accuracy of the RV results after excluding poor-quality fits.  We find that the systems most amenable to successful recovery are those with an orbital period shorter than 100 days and planetary mass greater than half a Jupiter mass (see Figure \ref{fig:Mp}).

We found that orbital periods as long as $\sim$180 days can be determined to a fractional accuracy of 10-20\%,
in which case around half of the orbital phase was observed, and the RV data usually show some linear trends. These systems usually contain a planet with a high mass yielding a large semi-amplitude. Detection of such systems in a campaign as mapped out here would prompt observers to take additional RV observations to refine the period.

\section{Discussion}
\label{sec:disc}

In this work, we have updated the simulation process introduced in Y19 by incorporating orbits with realistic eccentricity distributions, and improved the recovery rates by using $P_{\rm cal}$ to refine the searched periods. The results from this work use more realistic descriptions of planetary orbits, and should be more useful for the observers to schedule photometric follow-up. If an observer has a particular single transit TESS candidate, they can calculate $P_{\rm cal}$, use the KELT light curve to compute $P_{\rm BLS}$ and the associated SPN, and check the confidence of $P_{\rm BLS}$ from Figure \ref{fig:spn}.  They can use that information to prioritize and schedule photometric observations during the predicted transit window for that candidate to confirm the ephemeris. A similar procedure can be used with any archival photometry, so long as a sensitivity analysis along the lines of Y19 and \ref{sec:precovery} is conducted to determine the SPN threshold associated with a given confidence level for a particular range of transit depth and duration.

That analysis can still be improved.  We assume central transits in our simulations, but real transits, with a range of impact parameters, will be shorter for the same stellar properties and orbital period, leading to additional errors in the value of $P_{\rm cal}$. However, the overall size of that effect is small \citep{Seager:2003}.  It is the case that one could estimate the impact parameter and other orbital properties from the \tess light curve, but we have assumed here that while the \tess light curve is likely to have high enough photometric precision to obtain fairly reliable measurements of the transit depth and duration, it should not be assumed that the transit shape, along with ingress/egress times, can always be measured to high precision. In cases with extremely high photometric precision, the transit shape itself can be used to identify or discount false positive scenarios, but that typically can only be done for very bright stars, such as the case of HR 858, which has multiple planets orbiting a $T=5.9$ star \citep{Vanderburg:2019}.  Another issue we did not consider is whether the presence of additional planets or stellar variability due to rotational modulation from spots/plages in the system could create RV signatures that would interfere with the ability to recover the orbit of the transiting planet.

We also explored in \S \ref{sec:EBs} how RV observations can be used for initial vetting of single-transit candidates, by simulating the use of two RV observations tied to the predicted quadrature times based on simplified orbital assumptions.  We showed that for most system configurations, such observations should be readily able to distinguish stellar binaries from planetary systems.

\begin{figure}[t]
\begin{center}
\makebox[\textwidth][c]{\includegraphics[width=1.0\textwidth]{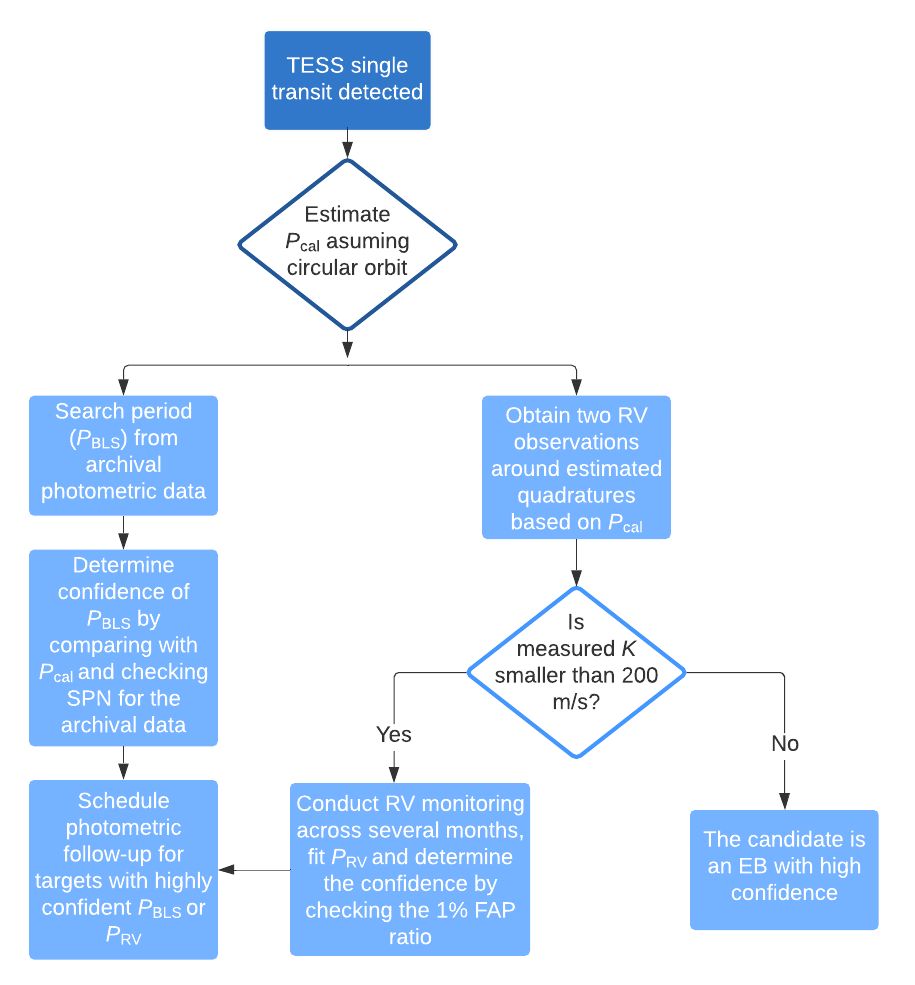}}
\caption{A flow chart illustrating the process to determine the ephemerides of TESS single transits described in this paper.}
\label{fig:flowchart}
\end{center}
\end{figure}

We then examined an example campaign of RV observations to confirm the single-transit candidates by measuring the orbital motion of the stellar host due to the planet.  We posited a particular observing campaign that is within the capability of multiple current facilities. For a TESS single transit, even after an initial pair of RV observations discounts the presence of an unblended stellar binary, there is a broad range of possible periods (Figure \ref{fig:hist}) even after accounting for the transit duration and stellar mass.  Our simulations show that RV campaigns of the type we considered can successfully measure the orbital motion due to the planet in $\sim$ 30\% of cases we simulated.
The results of such an RV campaign, making use of modest telescope and spectroscopic facilities, can constrain orbital period to within 5\% with 85\% confidence and the mass of the companion to within 50\% with 88\% confidence.  
That information can then be used to plan a photometric campaign over one or several nights to catch another transit and precisely determine the ephemeris for future atmospheric study, or conduct additional, targeted RV observations, potentially with higher RV precision to further constrain the period before conducting photometric observations.  A flow chart in Figure \ref{fig:flowchart} provides a schematic illustration of this RV follow-up process we have outlined.  While this approach requires a number of steps to refine the precise orbital period, it will typically be more feasible for most observers than blind photometric follow-up of a candidate, as in the case of NGTS-11b described below.

The achievable precision of an actual RV follow-up campaign depends on many factors, such as weather, observational windows, the number of RV observations that can be acquired, the properties of the target stars (e.g., magnitude, rotational velocity, effective temperature), the resolution of the spectrograph ($\frac{\lambda}{\Delta \lambda}$), the exposure time, etc. Therefore, we find it impractical to incorporate all these factors into a generalized simulation, but these results can be used as a set of guidelines for real-world follow-up efforts. In future work, we intend to explore how the fitting of the RV signals can be improved with other software tools such as EXOFASTv2 \citep{Eastman:2019} or Juliet \citep{Espinoza:2019}, as opposed to RadVel.

The TESS Single Transit Planet Candidates (TSTPC) working group has identified a sample of more than 100 single transit candidates in TESS FFI data during Sectors 1 through 13 (Villanueva et al in prep).  Assuming circular orbits to calculate the orbital period, and estimating the planetary mass using the calculated planetary radius and the mass-radius relation from \citet{Chen:2017} we find that about 70\% of the TSTPC-identified candidates would be amenable to confirmation using these techniques.

\citet{Cooke:2020} have approached follow-up observations of TESS single transits in a different way, as mentioned in section \ref{sec:intro}. That work asks a narrower question than we deal with here.  They investigate how one should conduct follow-up observations when deciding between a blind photometric search or RV search, for two (or three) specific observing facilities.  They calculate the  effectiveness of using photometry or spectroscopy for exoplanet follow-up as a function of $\rm R_{\star}$, $\rm R_p$ and $\rm P$. They assume circular orbits for the transit candidates, and expect full, dedicated access the the observing facilities, which in that case consist of the NGTS photometric telescopes, and the CORALIE and HARPS spectrographs.  For photometric follow-up observations, they model regular, nightly observations, in a mode exemplified by the successful confirmation of the ephemeris of the planet NGTS-11b.  In that case, \citet{Gill:2020b} describe how the follow-up procedure dedicated 79 nights of photometric monitoring in a blind search for the detection of a second transit of a TESS single-transit candidate.  Such an approach is complementary to the process we describe in this paper and in Y19.  When archival photometry is available, we can restrict the likely ephemerides to certain time ranges or to exact values, but if such photometry is not available, the approach used by \citet{Cooke:2020} provides an alternate path, when sufficient resources are available.  The RV follow-up approach they describe accounts for a more exact calculation of the RV precision as a function of target magnitude for NGTS,
HARPS, and CORALIE and the associated SNR of the RV detection, and the assumption of circular orbits rather than a range of eccentricities represents a different set of assumptions than we use.  We believe that the approach described in this work should be useful for observers working with a range of follow-up observing facilities and various levels of access, while the \citet{Cooke:2020} analysis is most appropriate for those with dedicated access to high-performance instruments. 

\section{Summary}
\label{sec:sum}

TESS will discover hundreds of planet candidates with long orbital periods via single transits in their TESS light curves during the prime mission \citep{Cooke:2018, Villanueva:2019}. Y19 demonstrated that KELT data can recover the ephemerides of some single-transit candidates to a fractional precision of 0.01\%.  In this work, we have incorporated more realistic models for TESS single transits with eccentric orbits instead of circular orbits, which is common for long-period exoplanets. We also improved the reliability of precovery using transit duration constraints.

In addition to improving the precovery simulations, we explored the use of RV follow-up observations in the case of single transit events. We found that the use of the estimated period based on a circular orbit to schedule reconnaissance RV observations around quadrature can efficiently distinguish EBs from planets.  For candidates that pass reconnaissance RV observations, we simulated RV monitoring campaigns to obtain an orbital solution.  We found that the use of a $\chi_{\rm dof}^{2}$ fit to a Keplerian model, combined with an FAP analysis, is sufficient to obtain an approximate orbital solution for planets more massive than $0.5 M_{\rm J}$ with orbital periods as long as 100 days, providing sufficient constraints for additional detailed orbital refinement and photometric ephemeris determination.

\acknowledgments{

JP and XY acknowledge support from NASA grant 80NSSC19K0387 under the TESS Guest Investigator program G011229. PD acknowledges support from a National Science Foundation (NSF) Astronomy \& Astrophysics Postdoctoral Fellowship under award AST-1903811. Part of this research was carried out at the Jet Propulsion Laboratory, California Institute of Technology, under a contract with the National Aeronautics and Space Administration (NASA). KGS acknowledges partial support from NASA grant 17-XRP17 2-0024. BSG was supported by a Thomas Jefferson Chair for Space Exploration endowment at the Ohio State University.  DJS is supported as an Eberly Research Fellow by the Eberly College of Science at the Pennsylvania State University. The Center for Exoplanets and Habitable Worlds is supported by the Pennsylvania State University, the Eberly College of Science, and the Pennsylvania Space Grant Consortium.

KELT was partially supported by NSF CAREER Grant AST-1056524. Funding for the \tess mission is provided by the NASA Explorer Program.  This work has made use of NASA's Astrophysics Data System.  This research made use of Astropy,\footnote{http://www.astropy.org} a community-developed core Python package for Astronomy \citep{astropy:2013, astropy:2018}.  We acknowledge support for the KELT project through the Vanderbilt Initiative in Data-intensive Astrophysics, Ohio State University, and Lehigh University. 

}

\bibliographystyle{apj}
\bibliography{main.bbl}

\end{document}